\documentclass[conference]{IEEEtran}

\ifCLASSINFOpdf
 
\else
  
\fi

\usepackage{amsmath,amssymb}
\usepackage{color}
\usepackage[noadjust]{cite}
\usepackage{graphicx}
\usepackage{epsfig}
\usepackage{breqn}
\usepackage[caption=false,font=footnotesize]{subfig}
\usepackage{array}

\newtheorem{theorem}{Theorem}[section]

\newtheorem{conjecture}[theorem]{Conjecture}

\allowdisplaybreaks

\newcommand\given[1][]{\:#1\vert\:}

\begin{document}

\title{Minimum node degree in inhomogeneous random key graphs with unreliable links}

\author{\IEEEauthorblockN{Rashad Eletreby and Osman Ya\u{g}an}
\IEEEauthorblockA{Department
of Electrical and Computer Engineering and CyLab, \\
Carnegie Mellon University, Pittsburgh,
PA, 15213 USA\\
reletreby@cmu.edu, oyagan@ece.cmu.edu}}

\maketitle

\begin{abstract}
We consider wireless sensor networks under a heterogeneous random key predistribution scheme and an on-off channel model. 
The heterogeneous key predistribution scheme has recently been introduced by Ya\u{g}an - as an extension to the Eschenauer and Gligor scheme - for the cases when the network consists of sensor nodes with varying level of resources and/or connectivity requirements, e.g., regular nodes vs. cluster heads. The network is modeled
by the intersection of the inhomogeneous random key graph (induced by the heterogeneous scheme) with an Erd\H{o}s-R\'enyi graph (induced by the on/off channel model).
We present conditions (in the form of zero-one laws) on how to scale the parameters of the intersection model so that with high probability all of its nodes are connected to at least $k$ other nodes; i.e., the minimum node degree of the graph is no less than $k$. We also present numerical results to support our results in the finite-node regime. The numerical results suggest that the conditions that ensure $k$-connectivity coincide with those ensuring the minimum node degree being no less than $k$.
\end{abstract}


\IEEEpeerreviewmaketitle

\section{Introduction}
{Wireless} sensor networks (WSNs) comprise low-cost, low-power, small sensor nodes that are typically deployed in hostile environments \cite{Akyildiz_2002}. WSNs have numerous applications including military, health, and environmental monitoring, among others \cite{Akyildiz_2002}. Ensuring secure communications in WSNs is expected to be a key 
challenge \cite{security_survey} given several limiting factors; e.g., limited computational capabilities, limited transmission power, lack of a priori knowledge of deployment configuration, and vulnerability to node capture attacks.
Random key predistribution schemes have been proposed with these limitations in mind and they have become widely regarded as the most feasible solutions in securing WSNs; e.g., see \cite[Chapter~13]{Raghavendra_2004} and \cite{camtepe_2005}, and references therein.

Random key predistribution schemes were first proposed by Eschenauer and Gligor   \cite{Gligor_2002}. 
Their scheme, hereafter referred to as the EG scheme,  operates as follows: prior to deployment, each sensor node is assigned a {\em random} set of  $K$ keys from a key pool of size $P$. Once deployed, two nodes can securely communicate only if they have at least one key in common. 
The EG scheme paved the way to several other variants, including the $q$-composite scheme  \cite{Haowen_2003}, and the random pairwise scheme \cite{Haowen_2003} among others.
Recently, Ya\u{g}an introduced a new variation of the EG scheme, referred to as the heterogeneous key predistribution scheme \cite{Yagan/Inhomogeneous}. The heterogeneous scheme addresses the case when the WSN comprises sensor nodes with varying level of resources and/or connectivity requirements, e.g., regular nodes vs. cluster heads, which is likely to hold for many WSN applications \cite{Yarvis_2005}. It is described as follows. 
Given $r$ classes, each sensor node is independently classified as a class-$i$ node with probability $\mu_i>0$ 
for each $i=1,\ldots, r$.
Then, sensors in class-$i$ are each given $K_i$ keys selected uniformly at random  from a key pool of size $P$. As before, sensors that share key(s) can communicate securely over an available channel after the deployment; see Section \ref{sec:Model} for details. 

In \cite{Yagan/Inhomogeneous}, Ya\u{g}an established critical conditions on the probability distribution $\pmb{\mu}=\{\mu_1,\mu_2,\ldots,\mu_r \}$, the key ring sizes $\pmb{K}=\{K_1,K_2,\ldots,K_r\}$, and the key pool size $P$ as a function of the network size $n$, so that the resulting WSN is securely connected with high probability. 
Although these results form a crucial starting point towards the analysis of feasibility of the heterogeneous key predistribution scheme, there remains to establish several important properties of the scheme to obtain a fuller understanding of its performance in securing WSNs. In particular,
 the connectivity results given in \cite{Yagan/Inhomogeneous} were obtained under the assumption of {\em full visibility}; i.e., that all pairs of sensors have a communication channel in between. This assumption is not likely to hold in practical WSNs  
given that the wireless medium of the communication is often unreliable and sensors typically have limited communication ranges. Moreover, the results in \cite{Yagan/Inhomogeneous} do not provide any information on the {\em reliability} of network connectivity against sensor or link failures. 
Reliability of WSNs is crucial since sensors may fail due to battery depletion or may get captured by an adversary, and wireless links may fail due to harsh environmental conditions.

Our paper completes a crucial first step towards addressing both of the aforementioned limitations of the results by Ya\u{g}an in \cite{Yagan/Inhomogeneous}. First of all, we consider the heterogeneous key predistribution scheme under {\em non-full visibility}. In particular, we consider an on/off communication model consisting of independent wireless channels each of which is either on (with probability $\alpha$), or off (with probability $1-\alpha$). Secondly, we focus on the $k$-connectivity property which implies that the network connectivity is preserved despite the failure of any $(k-1)$ nodes or links \cite{PenroseBook}. Accordingly, $k$-connectivity provides a guarantee of network reliability against the potential failures of sensors or links. 

Our approach is based on modeling the WSN by an appropriate random graph and then establishing scaling conditions on the model parameters such that  certain desired properties hold with high probability (whp) as the number of nodes $n$ gets large. 
The heterogeneous key predistribution scheme induces the inhomogeneous random key graph \cite{Yagan/Inhomogeneous}, denoted hereafter by $\mathbb{K}(n,\pmb{\mu},\pmb{K},P)$, while the on/off communication model leads to the standard Erd\H{o}s-R\'enyi (ER) graph \cite{ER}, denoted by $\mathbb{G}(n,\alpha)$. Hence, the appropriate random graph model here is the intersection of the inhomogeneous random key graph with ER graph, denoted $\mathbb{K \cap G}(n;\pmb{\mu},\pmb{K},P,\alpha)$. 

Our main result (see Theorem \ref{theorem:min_node_degree}) is a zero-one law for the minimum node degree of $\mathbb{K \cap G}(n;\pmb{\mu},\pmb{K},P,\alpha)$ to be no less than $k$ for any non-negative integer $k$. More precisely, we present conditions  on how to scale the parameters of $\mathbb{K \cap G}(n;\pmb{\mu},\pmb{K},P,\alpha)$ so that its minimum node degree is no less than $k$ with high probability when the number of nodes $n$ gets large. The zero-law for minimum node degree being no less than $k$ already implies the zero-law for $k$-connectivity since a graph can not be $k$-connected unless all nodes have degree at least $k$. Furthermore, we show by simulations that minimum node degree being no less than $k$ and $k$-connectivity properties exhibit almost equal (empirical) probabilities, which prompts us to conjecture that the one-law of $k$-connectivity will also follow under the same critical scaling and conditions provided here;
see Section \ref{sec:numerical} for details. 

Throughout, an event is said to hold {\em with high
probability} (whp) if it holds with probability $1$ as $n \to
\infty$. 
For sequences $\{a_n\},\{b_n\}$,
we use
$a_n = o(b_n)$,  $a_n=\omega(b_n)$, and $a_n = O(b_n)$ with their meaning in
the standard Landau notation. 

\section{The Model}
\label{sec:Model}
We consider a network consisting of $n$ sensor nodes labeled as $v_1, v_2, \ldots,v_n$. Each sensor node is assigned to one of the possible $r$ classes, e.g., priority levels, according to a probability distribution $\pmb{\mu}=\{\mu_1,\mu_2,\ldots,\mu_r\}$ with $\mu_i >0$
for each  $i=1,\ldots,r$; clearly it is also needed that $\sum_{i=1}^r \mu_i=1$. Prior to deployment, each class-$i$ node is given $K_i$ cryptographic keys selected uniformly at random from a pool of size $P$. 
Hence, the key ring $\Sigma_x$ of node $v_x$ is a $\mathcal{P}_{K_{t_x}}$-valued random variable (rv) where $\mathcal{P}_A$ denotes the collection of all subsets of $\{1,\ldots,P\}$ with exactly $A$ elements and $t_x$ denotes the class of node $v_x$. The rvs $\Sigma_1, \Sigma_2, \ldots, \Sigma_n$ are then i.i.d. with
\begin{equation}
\mathbb{P}[\Sigma_x=S \mid t_x=i]= \binom P{K_i}^{-1}, \quad S \in \mathcal{P}_{K_i}.
\nonumber
\end{equation}
After the deployment, two sensors can communicate securely over an existing communication channel if they have at least one key in common.

Throughout, we let $\pmb{K}=\{K_1,K_2,\ldots,K_r\}$, and assume without loss of generality that $K_1 \leq K_2 \leq \ldots \leq K_r$. 
Consider a random graph $\mathbb{K}$ induced on the vertex set $\mathcal{V}=\{v_1,\ldots,v_n\}$ such that two distinct nodes $v_x$ and $v_y$ are adjacent in $\mathbb{K}$, denoted by the event $K_{xy}$, if they have at least one cryptographic key in common, i.e.,
\begin{equation}
K_{xy} :=\left[\Sigma_x \cap \Sigma_y \neq \emptyset\right].
\label{adjacency_condition}
\end{equation}
The adjacency condition (\ref{adjacency_condition}) characterizes the inhomogeneous random key graph  $\mathbb{K}(n;\pmb{\mu},\pmb{K},P)$ as introduced recently by Ya\u{g}an \cite{Yagan/Inhomogeneous}.
This model is also known in the literature  as
the {\em general random intersection graph}; e.g., see \cite{Zhao_2014,Rybarczyk,Godehardt_2003}.

The inhomogeneous random key graph models the {\em cryptographic} connectivity of the underlying WSN.
In particular, the probability that a class-$i$ node and a class-$j$ node are adjacent, denoted by $p_{ij}$ is given by
\begin{equation}
p_{ij}= \mathbb{P}[K_{xy}] = 1-{\binom {P-K_i}{K_j}}\Bigg/{\binom {P}{K_j}}
\label{eq:osy_edge_prob_type_ij}
\end{equation}
as long as $K_i + K_j \leq P$; otherwise if $K_i +K_j > P$, we clearly have $p_{ij}=1$.
We also find it useful define the \textit{mean} probability $\lambda_i$ of edge occurrence for a class-$i$ node in $\mathbb{K}(n;\pmb{\mu},\pmb{K},P)$. With arbitrary nodes $v_x$ and $v_y$, we have
\begin{align}
\lambda_i&=\mathbb{P}[K_{xy} \given[\big] t_x=i ] 
=\sum_{j=1}^r p_{ij} \mu_j,  \quad i=1,\ldots, r,
 \label{eq:osy_mean_edge_prob_in_RKG}
\end{align}
as we condition on the class $t_y$ of node $v_y$.

In  this work, we consider the communication topology of the WSN as consisting of independent 
channels that are either {\em on} (with probability $\alpha$) or {\em off} (with probability $1-\alpha$). 
More precisely, let $\{B_{ij}(\alpha), 1 \leq i < j \leq n\}$ denote i.i.d Bernoulli rvs, each with success probability $\alpha$. The communication channel between two distinct nodes $v_x$ and $v_y$ is on (respectively off) if $B_{xy}(\alpha)=1$ (respectively if $B_{xy}(\alpha)=0$). 
This simple on/off channel model enables a comprehensive analysis of the properties of interest of the resulting WSN, e.g., its connectivity, and was shown to be a good approximation to the more realistic disk model \cite{Gupta99,ZhaoAllerton2014} in many similar settings; e.g., see \cite{Yagan/EG_intersecting_ER,YaganPairwise,YaganICC2011}.
The on/off channel model induces  a standard Erd\H{o}s-R\'enyi graph $\mathbb{G}(n;\alpha)$  \cite{Bollobas}, defined  on the vertices $\mathcal{V}=\{v_1,\ldots,v_n\}$ such that 
  $v_x$ and $v_y$ are adjacent, denoted $C_{xy}$,  
if $B_{xy}(\alpha)=1$.


We model the overall topology of a WSN by the intersection of the inhomogeneous random key graph $\mathbb{K}(n;\pmb{\mu},\pmb{K},P)$ with the ER graph $\mathbb{G}(n;\alpha)$. Namely, nodes  $v_x$ and $v_y$ are adjacent in $\mathbb{K \cap G}(n;\pmb{\mu},\pmb{K},P,\alpha)$, denoted $E_{xy}$, if and only if they are adjacent in both $\mathbb{K}$ \textit{and} $\mathbb{G}$. In other words, the edges in the intersection graph
$\mathbb{K \cap G}(n;\pmb{\mu},\pmb{K},P,\alpha)$ represent pairs of sensors that
can securely communicate as they have i) a communication link available in between, and ii) shared cryptographic key(s).
Therefore, studying the connectivity properties of $\mathbb{K \cap G}(n;\pmb{\mu},\pmb{K},P,\alpha)$ amounts to studying the secure connectivity of heterogenous WSNs under the on/off channel model.


To simplify the notation, we let $\pmb{\theta}=(\pmb{\mu},\pmb{K},P)$, and $\pmb{\Theta}=(\pmb{\theta},\alpha)$. The probability of edge existence between a class-$i$ node $v_x$ and a class-$j$ node $v_y$ in $\mathbb{K \cap G}(n;\pmb{\Theta})$ is given by
\begin{equation} \nonumber
\mathbb{P}[E_{xy} \given[\Big] t_x=i,t_y=j]=\mathbb{P}[K_{xy} \cap C_{xy} \given[\big] t_x=i,t_y=j]=\alpha p_{ij}
\end{equation}
by independence. Similar to (\ref{eq:osy_mean_edge_prob_in_RKG}), the mean edge probability for a class-$i$ node in $\mathbb{K \cap G}(n;\pmb{\Theta})$, denoted by $\Lambda_i$, is given by
\begin{align} 
\Lambda_i = \sum_{j=1}^r \mu_j \alpha p_{ij} = \alpha \lambda_i, \quad i=1,\ldots, r.
\label{eq:osy_mean_edge_prob_in_system}
\end{align}

Throughout, we assume that the number of classes $r$ is fixed and does not scale with $n$, and so are the probabilities $\mu_1, \ldots,\mu_r$. All of the remaining parameters are assumed to be scaled with $n$.

\section{Main Results and Discussion}
\label{sec:results}
We refer to a mapping $K_1,\ldots,K_r,P: \mathbb{N}_0 \rightarrow \mathbb{N}_0^{r+1}$ as a \textit{scaling} (for the inhomogeneous random key graph) as long as the conditions
\begin{equation}
2 \leq K_{1,n} \leq K_{2,n} \leq \ldots \leq K_{r,n} \leq P_n/2
\label{scaling_condition_K}
\end{equation}
are satisfied  for all $n=2,3,\ldots$. Similarly any mapping $\alpha: \mathbb{N}_0 \rightarrow (0,1)$ defines a scaling for the ER graphs. As a result, a mapping $\pmb{\Theta} : \mathbb{N}_0 \rightarrow \mathbb{N}_0^{r+1} \times (0,1)$ defines a scaling for the intersection graph $\mathbb{K \cap G}(n;\pmb{\Theta}_n)$ given that condition (\ref{scaling_condition_K}) holds. We remark that under (\ref{scaling_condition_K}), the edge probabilities $p_{ij}$ will be given by
(\ref{eq:osy_edge_prob_type_ij}).

Our main result is a zero-one law for the minimum node degree being no less than $k$ in the inhomogeneous random key graph intersecting ER graph.

\begin{theorem}
\label{theorem:min_node_degree}
{\sl
Consider a probability distribution $\pmb{\mu}=(\mu_1,\mu_2,\ldots,\mu_r)$ with $\mu_i >0$ for $i=1,\ldots,r$, a scaling $K_1,\ldots,K_r,P: \mathbb{N}_0 \rightarrow \mathbb{N}_0^{r+1}$, and a scaling $\alpha: \mathbb{N}_0 \rightarrow (0,1)$. Let the sequence $\gamma: \mathbb{N}_0 \rightarrow \mathbb{R}$ be defined through 
\begin{equation}
\Lambda_1(n)=\alpha_n \lambda_1(n) = \frac{\log n + (k-1)\log \log n+\gamma_n}{n}, \label{scaling_condition_KG}
\end{equation}
for each $n=1,2, \ldots$. 

(a) If $\lambda_1(n)=o(1)$, we have
\begin{equation} \nonumber
\lim_{n \to \infty} \mathbb{P} \left[ \begin{split} &\text{ Minimum node degree} \text{ of }  \\ &\mathbb{K \cap G}(n;\pmb{\Theta}_n) \text{ is at least } k \end{split} \right]= 0   ~~~ \text{ if } \lim_{n \to \infty} \gamma_n=-\infty
\end{equation}

(b) We have
\begin{equation} \nonumber
\lim_{n \to \infty} \mathbb{P} \left[ \begin{split} &\text{ Minimum node degree of}\\ & 
 \mathbb{K \cap G}(n;\pmb{\Theta}_n) \text{ is at least }k \end{split} \right]=1  ~~~ \text{ if } \lim_{n \to \infty} \gamma_n=\infty.
\end{equation}

}
\end{theorem}

In words, Theorem~\ref{theorem:min_node_degree} states that the minimum node degree in $\mathbb{K \cap G}(n;\pmb{\Theta}_n)$ is greater than or equal $k$ whp if the mean degree of class-$1$ nodes (i.e., $n \Lambda_1(n)$) is scaled as $\left(\log n+(k-1) \log \log n+\gamma_n\right)$ for some sequence $\gamma_n$ satisfying $\lim_{n \to \infty} \gamma_n=\infty$. On the other hand, if the sequence $\gamma_n$ satisfies $\lim_{n \to \infty} \gamma_n=-\infty$, then whp $\mathbb{K \cap G}(n;\pmb{\Theta}_n)$ has at least one node with degree strictly less than $k$. This shows that the critical scaling for the minimum node degree of 
$\mathbb{K \cap G}(n;\pmb{\Theta}_n)$ being greater than or equal $k$
is given by $\Lambda_1(n)=\frac{\log n+(k-1)\log \log n}{n}$, with the sequence $\gamma_n:\mathbb{N}_0 \rightarrow \mathbb{R}$ measuring the deviation of $\Lambda_1(n)$ from the critical scaling.

The zero-one law established here for the minimum node degree
of $\mathbb{K \cap G}(n;\pmb{\Theta}_n)$ shall be regarded as a crucial
first step towards establishing a similar result on its $k$-connectivity; namely the property 
that the graph will remain connected despite the deletion of any $k-1$ nodes or edges.
In fact, Theorem~\ref{theorem:min_node_degree} 
already implies the zero-law for the $k$-connectivity, i.e., that
\[
\lim_{n \to \infty} \mathbb{P} \left[ \begin{split} & \mathbb{K \cap G}(n;\pmb{\mu},\pmb{\Theta}_n) \\ & \text{is $k$-connected} \end{split} \right]= 0   \quad \text{ if } \lim_{n \to \infty} \gamma_n=-\infty.
\]
This is because a graph can not be $k$-connected unless all its nodes have degree at least $k$. Also, 
for several classes of random graphs it is known that the conditions that ensure $k$-connectivity coincide with those ensuring minimum node degree to be at least $k$; e.g., 
random key graphs \cite{Jun/K-Connectivity}, ER graphs \cite{Bollobas}, and random geometric key graphs \cite{PenroseBook}. Here we confirm this via numerical simulations (see Section~\ref{sec:numerical}) for the intersection of inhomogeneous random key graph and ER graph, prompting us to  cast the following conjecture.

\begin{conjecture}
{\sl
Consider a probability distribution $\pmb{\mu}=(\mu_1,\mu_2,\ldots,\mu_r)$ with $\mu_i >0$ for $i=1,\ldots,r$, a scaling $K_1,\ldots,K_r,P: \mathbb{N}_0 \rightarrow \mathbb{N}_0^{r+1}$, and a scaling $\alpha: \mathbb{N}_0 \rightarrow (0,1)$. Let the sequence $\gamma: \mathbb{N}_0 \rightarrow \mathbb{R}$ be defined through 
(\ref{scaling_condition_KG}). 
With $\lambda_1(n)=o(1)$ and possibly under some additional conditions, we have
\begin{equation} \nonumber
\lim_{n \to \infty} \mathbb{P} \left[ \begin{split} & \mathbb{K \cap G}(n;\pmb{\Theta}_n) \\ & \text{is $k$-connected} \end{split} \right]=
\left \{
\begin{array}{cl}
0     &  \text{ if ~} \lim_{n \to \infty} \gamma_n=-\infty   \\
  1   &    \text{ if ~} \lim_{n \to \infty} \gamma_n= +\infty
\end{array}
 \right.
\end{equation}
}
\end{conjecture}

We comment on the additional technical condition $\lambda_1(n)=o(1)$. This is enforced here mainly for technical reasons for the proof of the zero-law of Theorem~\ref{theorem:min_node_degree} to work. A similar condition on the edge probability of the (homogeneous) random key graph was required in \cite{Jun/K-Connectivity} for obtaining the zero-law for the minimum node degree being no less than $k$ in the (homogeneous) random key graph intersecting ER graph. In view of \cite[Lemma 4.2]{Yagan/Inhomogeneous}, this condition is equivalent to
\[
K_{\textrm{min}} K_{\textrm{avg}} = o(P_n),
\]
with $K_{\textrm{min}}$ and $K_{\textrm{avg}}$  denoting the minimum and mean key ring sizes, respectively. Considering the fact that 
key pool size $P$ is expected to be orders of magnitude larger than any key ring size in the network \cite{Yagan/Inhomogeneous} this condition would be satisfied in most WSN implementations.

In comparison with the existing literature on similar models, our result can be seen to extend the work by Zhao et al. \cite{Jun/K-Connectivity} on the homogeneous random key graph intersecting ER graph to the heterogeneous setting. There, zero-one laws for the property that the minimum node degree is no less than $k$ and the property that the graph is $k$-connected were established for $\mathbb{K \cap G}(n,K,P,\alpha_n)$. With $r=1$, i.e., when all nodes belong to the same class and thus receive the same number $K$ of keys,  Theorem~\ref{theorem:min_node_degree}  recovers the result of Zhao et al. for the property that the minimum node degree is no less than $k$ (See \cite[Theorem~1]{Jun/K-Connectivity}).

Our paper can also be seen as an extension of the work by Ya\u{g}an \cite{Yagan/Inhomogeneous} who  considered the inhomogeneous random key graph $\mathbb{K}(n,\pmb{\mu},\pmb{K},P)$ under {\em full} visibility; i.e., when all pairs of nodes have a communication channel in between. There,  Ya\u{g}an established zero-one laws for the absence of isolated nodes (i.e., absence of nodes with degree zero) and $1$-connectivity. Our work generalizes Ya\u{g}an's results on two levels. Firstly, we consider more practical WSN scenarios where the unreliability of wireless communication channels are taken into account through the on/off channel model. Secondly, in addition to the property that the graph has no isolated nodes (i.e., the minimum node degree is no less than $1$), we also establish results for a general minimum node degree $k=0,1,\ldots$.

Finally,  our work (with $\alpha_n=1$ for each $n=2, 3, \ldots$) improves upon the results by Zhao et al. \cite{Zhao_2014} for the 
minimum node degree in the inhomogeneous random key graph; therein, this model was referred to as the general random intersection graph. Even though Zhao et al. \cite{Zhao_2014} considered strong graph properties 
including $k$-connectivity and $k$-robustness, the additional conditions required by their main result renders it inapplicable in practical WSN implementations.
This issue is discussed at length in \cite[Section 3.3]{Yagan/Inhomogeneous}, but we give a summary here for completeness.
With $X_n$ denoting the random variable representing the number of keys assigned to an arbitrary node in the network, the main result in  \cite{Zhao_2014} requires
$
\text{Var} [X_n]=o \left( \frac{\left( \mathbb{E}[X] \right)^2}{n \left( \log n \right)^2} \right)
$,
which implies that 
\begin{equation}
\mathbb{E}[X_n]=\omega(\sqrt{n} \log n).
\label{eq:review:2}
\end{equation}
Particularly, the results in \cite{Zhao_2014} hold only if the mean number of keys assigned to a sensor node is much larger than $\sqrt{n} \log n$.
However, in most WSN applications, sensor nodes will have very limited memory and computational capabilities \cite{Akyildiz_2002}. Therefore, the condition (\ref{eq:review:2}) is not likely to hold in {\em large} WSNs. Instead, the only condition required by our main result can be seen to be $\mathbb{E}[X_n] = o(\sqrt{P_n})$.


%

\section{Numerical Results}
\label{sec:numerical}
We now present numerical results to support Theorem~\ref{theorem:min_node_degree} in the finite node regime. In all experiments, we fix the number of nodes at $n = 500$ and the size of the key pool at $P = 10^4$. For better visualization, we use the curve fitting tool of MATLAB.

\begin{figure}[t]
\vspace{-4mm}
\centerline{\includegraphics[scale=0.25]{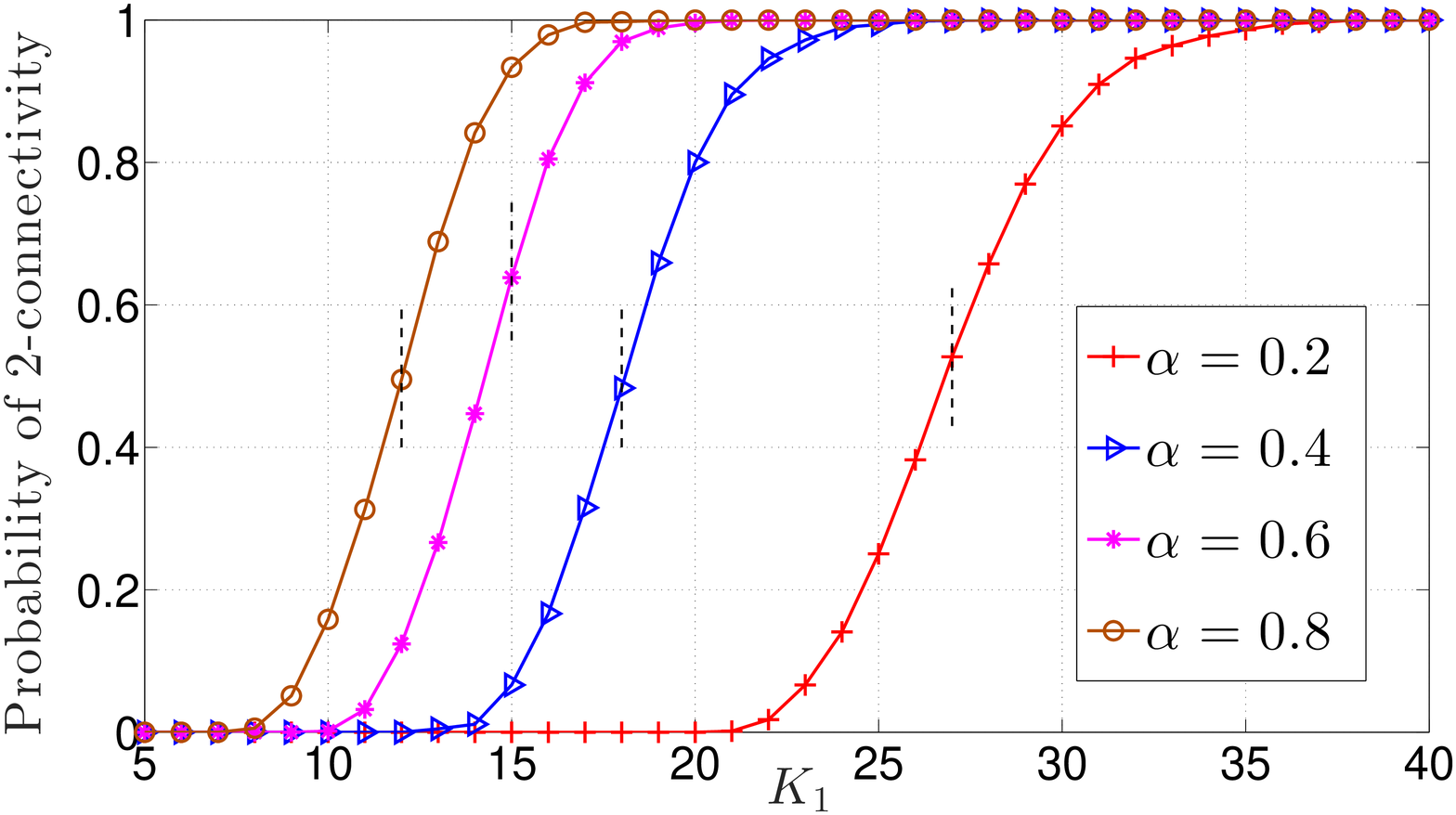}}
\vspace{-3mm}
\caption{Empirical probability that $\mathbb{K \cap G}(n;\pmb{\theta},\alpha)$ is $2$-connected as a function of $\pmb{K}$ for $\alpha = 0.2$, $\alpha = 0.4$, $\alpha = 0.6$, $\alpha = 0.8$ with $n = 500$ and $P = 10^4$. In each case, the empirical probability value is obtained by averaging over $200$ experiments. Vertical dashed lines stand for the critical threshold of $2$-connectivity asserted by Theorem~\ref{theorem:min_node_degree}.}
\label{fig:1}
\vspace{-4mm}
\end{figure}

In Figure~\ref{fig:1}, we consider the channel parameters $\alpha = 0.2$, $\alpha = 0.4$, $\alpha = 0.6$, and $\alpha = 0.8$, while varying the parameter $K_1$, i.e., the smallest key ring size, from $5$ to $40$. The number of classes is fixed to $2$, with $\pmb{\mu}=\{0.5,0.5\}$. For each value of $K_1$, we set $K_2=K_1+10$. For each parameter pair $(\pmb{K}, \alpha)$, we generate $200$ independent samples of the graph $\mathbb{K \cap G}(n;\pmb{\theta},\alpha)$ and count the number of times (out of a possible 200) that the obtained graphs i) have minimum node degree no less than $2$ and ii) are $2$-connected. Dividing the counts by $200$, we obtain the (empirical) probabilities for the events of interest.  In all cases considered here, we observe that $\mathbb{K \cap G}(n;\pmb{\theta},\alpha)$ is $2$-connected whenever it has minimum node degree no less than $2$ yielding the same empirical probability for both events. This supports our conjecture that the 
properties of $k$-connectivity and minimum node degree being larger than $k$
are asymptotically equivalent in $\mathbb{K \cap G}(n;\pmb{\theta},\alpha)$.

For each $\alpha$ value, we show the critical threshold of $k$-connectivity \lq\lq predicted" by Theorem~\ref{theorem:min_node_degree} by a vertical dashed line. More specifically, the vertical dashed lines stand for the minimum integer value of $K_1$ that satisfies
\begin{equation}
\lambda_1(n)=\sum_{j=1}^2 \mu_j \left( 1- \frac{\binom{P-K_j}{K_1}}{\binom{P}{K_1}} \right) > \frac{1}{\alpha} \frac{\log n+\log \log n}{n}.
\label{eq:numerical_critical}
\end{equation}
We see from Figure~\ref{fig:1} that the probability of $k$-connectivity transitions from zero to one within relatively small variations in $K_1$. Moreover, the critical values of $K_1$ obtained by (\ref{eq:numerical_critical}) lie within the transition interval.  

\begin{figure}[t]
\vspace{-0.8mm}
\centerline{\includegraphics[scale=0.25]{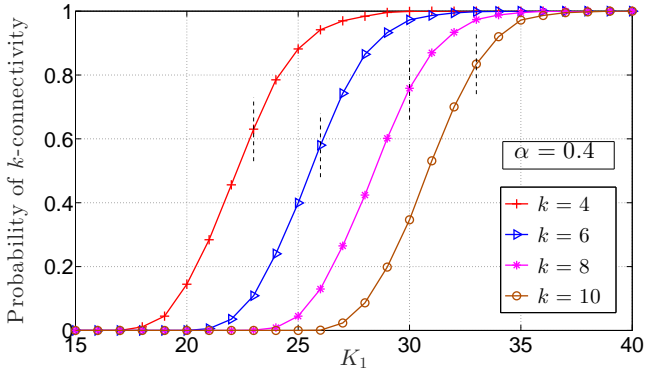}}
\vspace{-3mm}
\caption{Empirical probability that $\mathbb{K \cap G}(n;\pmb{\theta},\alpha)$ is $k$-connected as a function of $K_1$ for $k=4$, $k=6$, $k=8$, and $k=10$, with $n = 500$ and $P = 10^4$. In each case, the empirical probability value is obtained by averaging over $200$ experiments. Vertical dashed lines stand for the critical threshold of $k$-connectivity asserted by Theorem~\ref{theorem:min_node_degree}.}
\label{fig:3}
\vspace{-4mm}
\end{figure}

In Figure~\ref{fig:3}, we consider four different values for $k$, namely we set $k = 4$, $k = 6$, $k = 8$, and $k=10$ while varying $K_1$ from $15$ to $40$. The number of classes is fixed to $2$ with $\pmb{\mu}=\{0.5,0.5\}$ and we set $K_2=K_1+10$ for each value of $K_1$. Using the same procedure that produced Figure~\ref{fig:1}, we obtain the empirical probability that $\mathbb{K \cap G}(n;\pmb{\theta},\alpha)$ is $k$-connected versus $K_1$. The critical threshold of $k$-connectivity asserted by Theorem~\ref{theorem:min_node_degree} is shown by a vertical dashed line in each curve.

\vspace{-0.9mm}
\section{Proof of Theorem~\ref{theorem:min_node_degree}}
The proof of Theorem~\ref{theorem:min_node_degree} is lengthy and technically involved. Therefore, we omit the details in this conference version. All details can be found in \cite{Rashad/kconnectivity}. 
In this section, we present an outline of our proof. The proof of Theorem~\ref{theorem:min_node_degree} relies on the methods of first and second moments (See \cite[p.~55]{JansonLuczakRucinski}) applied to the number of nodes with degree $\ell$ in $\mathbb{K \cap G}(n;\pmb{\Theta}_n)$. Let $X_{\ell}(n;\pmb{\Theta}_n)$ denote the total number of nodes with degree $\ell$ in $\mathbb{K \cap G}(n;\pmb{\Theta}_n)$, namely,
\begin{equation}
X_{\ell}(n;\pmb{\Theta}_n)=\sum_{i=1}^n \pmb{1}\left[D_{i,\ell}(n;\pmb{\Theta}_n) \right]
\label{eq:degree_3}
\end{equation}
where $D_{i,\ell}(n;\pmb{\Theta}_n)$ denotes the event that node $v_i$ in $\mathbb{K \cap G}(n;\pmb{\Theta}_n)$ has degree $\ell$ for each $i=1,2,\ldots,n$. Throughout, we simplify the notation by writing $D_{i,\ell}$ instead of $D_{i,\ell}(n;\pmb{\Theta}_n)$. 

By using the methods of first and second moments we obtain
\begin{equation}
\mathbb{P}\left[X_{\ell}(n;\pmb{\Theta}_n)=0\right] \leq 1-\frac{\mathbb{E}\left[X_{\ell}(n;\pmb{\Theta}_n)\right]^2}{\mathbb{E}\left[X_{\ell}(n;\pmb{\Theta}_n)^2\right]},
\label{eq:degree_1}
\end{equation}
and
\begin{equation}
\mathbb{P}\left[X_{\ell}(n;\pmb{\Theta}_n)=0\right] \geq 1-\mathbb{E}\left[X_{\ell}(n;\pmb{\Theta}_n)\right] 
\label{eq:degree_2}
\end{equation}

Recalling (\ref{eq:degree_3}), we get
\begin{align} 
\mathbb{E}\left[X_{\ell}(n;\pmb{\Theta}_n)\right] &=\sum_{i=1}^n \mathbb{P}\left[D_{i,\ell} \right] =n \mathbb{P}\left[D_{x,\ell}\right] 
\label{eq:degree_4}
\end{align}
upon using the exchangeability of the indicator rvs $\left\{ \pmb{1}\left[ D_{i,\ell} \right];i=1,\ldots,n \right\}$. Similarly, it is easy to see that
\begin{equation}
\mathbb{E}\left[ \left( X_{\ell}(n;\pmb{\Theta}_n) \right)^2 \right]=n \mathbb{P}\left[ D_{x,\ell} \right]+n(n-1) \mathbb{P}\left[ D_{x,\ell} \cap D_{y,\ell} \right]
\label{eq:degree_5}
\end{equation}

Combining (\ref{eq:degree_4}) and (\ref{eq:degree_5}), we obtain
\begin{align}
\frac{\mathbb{E}\left[X_{\ell}(n;\pmb{\Theta}_n)^2\right]}{\mathbb{E}\left[X_{\ell}(n;\pmb{\Theta}_n)\right]^2} =&\frac{1}{n \mathbb{P}\left[ D_{x,\ell} \right]}
+ \left(\frac{n-1}{n}\right) \frac{\mathbb{P}\left[ D_{x,\ell} \cap D_{y,\ell} \right]}{\left\{ \mathbb{P}\left[ D_{x,\ell} \right]\right\}^2}.
\label{eq:degree_6}
\end{align}

The one-law states that the minimum node degree in $\mathbb{K \cap G}(n;\pmb{\Theta}_n)$ is no less than $k$ asymptotically almost surely (a.a.s.). Put differently, the one-law requires that the graph has no nodes with degree $\ell =0, 1, \ldots, k-1$, i.e.,
\begin{equation}
\lim_{n \to \infty} \mathbb{P}\left[X_{\ell}(n;\pmb{\Theta}_n)=0\right]=1, \quad  \ell =0,1, \ldots, k-1,
\label{eq:degree_7}
\end{equation}
In view of (\ref{eq:degree_2}) and (\ref{eq:degree_4}), we see that (\ref{eq:degree_7}) and hence the one-law
would follow upon showing
\begin{equation}
\lim_{n \to \infty} n \mathbb{P}\left[D_{x,\ell}\right]=0, \quad  \ell = 0, 1, \ldots, k-1.
\label{eq:to_show1}
\end{equation}

For the zero-law, we need to show that there exists $\ell =0,1, \ldots, k-1$ such that
\begin{equation}
\lim_{n \to \infty} \mathbb{P}\left[X_{\ell}(n;\pmb{\Theta}_n)=0\right]=0.
\label{eq:degree_9}
\end{equation}
In other words, we need to show that there is at least one node with degree $\ell = 0, 1, \ldots, k-1$, so that
the minimum node degree in $\mathbb{K \cap G}(n;\pmb{\Theta}_n)$
is less than $k$ a.a.s.

In view of (\ref{eq:degree_1}) and (\ref{eq:degree_6}), (\ref{eq:degree_9}) will follow if we show that
\begin{equation} \label{eq:to_show2}
\lim_{n \to \infty} n \mathbb{P}\left[D_{x,\ell}\right]=\infty,
\end{equation}
and
\begin{equation} \label{eq:to_show3}
\mathbb{P}\left[ D_{x,\ell} \cap D_{y,\ell} \right] \sim \left\{\mathbb{P}\left[ D_{x,\ell} \right]\right\}^2,
\end{equation}
for some $\ell=0,1,\ldots,k-1$ under the enforced assumptions. 

Collecting, the proof of Theorem~\ref{theorem:min_node_degree} passes through establishing
(\ref{eq:to_show1}), (\ref{eq:to_show2}), and (\ref{eq:to_show3}) under appropriate conditions. 
Due to space limitations, details are given in  \cite{Rashad/kconnectivity}.

%

\vspace{-1.5mm}
\section*{Acknowledgment}
This work has been supported by the Department of Electrical and Computer Engineering at Carnegie Mellon University and by a gift from Persistent Systems, Inc.. O. Ya\u{g}an acknowledges the Berkman Faculty Development Grant and R. Eletreby acknowledges William J. Happel Fellowship.
\ifCLASSOPTIONcaptionsoff
  \newpage
\fi

\vspace{-1mm}
\bibliographystyle{IEEEtran}
\bibliography{IEEEabrv,ISIT}

\end{document}